\documentclass[prd,aps,showpacs,superscriptaddress,twocolumn]{revtex4}
\usepackage{graphicx}

\begin{document}

%%\begin{flushright}
%%Liverpool preprint  LTH 696\\
%%Swansea preprint: SWAT/06/460\\
%%Tor Vergata preprint: ROM2F/2006/09
%% \end{flushright}

%%\draft  % \draft command makes pacs numbers print

\title{Decay constants of P-wave heavy-light mesons from
unquenched lattice QCD}

\author{G.Herdoiza}
\affiliation{
Department of Physics,
University of Wales Swansea,
Singleton Park,
Swansea, SA2 8PP, UK
\\ and \\
INFN, Sezione di Roma Tor Vergata,
Via della Ricerca Scientifica, 1
I-00133 Rome, Italy
 }
\email{Gregorio.Herdoiza@roma2.infn.it}
\author{C.~McNeile}
\email{mcneile@amtp.liv.ac.uk}
\author{C.~Michael}
\email{cmi@liverpool.ac.uk}
\collaboration{UKQCD Collaboration}
\affiliation{
Theoretical Physics Division, Dept. of Mathematical Sciences,
          University of Liverpool, Liverpool L69 3BX, UK}

\begin{abstract}
 We review some decays that require knowledge of the decay constants of
$0^{+}$ heavy-light mesons. We compute the decay constants of  P-wave
heavy-light mesons from unquenched lattice QCD, 
with two degenerate flavours of sea quarks, 
at a single lattice
spacing.
The lightest
sea quark mass used in the calculation is a third of the 
strange quark mass.
For the charm-strange meson we obtain the 
decay constant: $f_{D_{s\,0^+}} =
340(110)$ MeV using our normalisation conventions.
We obtain the
$f_{P_s}^{static}$ (static-strange P-wave) decay constant as $302(39)$
MeV. 

 \end{abstract}

\maketitle
% body of paper here

\section{Introduction}

%% why the decay constants are useful
%%
%%  fd_s
%%  CLEO-c results
%%  latest HPQCD/FNAL results
%%

The decay constants of heavy-light P-wave
mesons have a number of important uses
in phenomenology~\cite{Colangelo:1991ug,Colangelo:1992kc}.
In this paper we use unquenched lattice QCD to compute
the decay constants of $0^+$ heavy-light mesons,
using heavy quarks with masses around that of the charm quark,
and with heavy quarks in the static limit.

Due to heavy-quark symmetry, there 
are four P-wave ($L=1$) heavy-light 
excited mesons,  noted for short $D^{\ast\ast}$,
organised in two doublets, one doublet
carrying a total angular momentum $j=1/2$ and the other $j=3/2$ for the 
light quarks. In the charm sector, we use the notation
that the scalar $J^P=0^+ (j=1/2)$ 
meson is $D_{0^+}$ when the light quark is an up or a down quark and 
$D_{s\,0^+}$ when it is a strange quark. In the static limit of HQET, the 
members of each of the two doublets are degenerate in mass.

 The low mass of the $D_s(2317)$ meson, recently discovered by 
BaBar~\cite{Aubert:2003fg} and confirmed by other experiments,
relative to the quark model predictions of Godfrey and
Isgur~\cite{Godfrey:1985xj,Godfrey:1986wj} was originally a puzzle.  
The quantum numbers of $D_s(2317)$ are consistent with
$J^P$ = $0^+$, but this needs confirmation~\cite{Eidelman:2004wy}.
There are some
speculations that this state may be a 
molecule~\cite{Barnes:2003dj}.  The theoretical
studies to understand the $D_s(2317)$ have recently been reviewed by
Colangelo et al.~\cite{Colangelo:2004vu}
and by Swanson~\cite{Swanson:2006st}.
The lattice QCD results for
the mass of the $D_{s\,0^+}$  state were tantalising close (within large
errors) to the experimental mass of the $D_s(2317)$
state~\cite{Dougall:2003hv,Ohta:2005cn}. However, the lattice
calculation of the $D_{s\,0^+}$ meson 
in~\cite{Bali:2003jv} did not agree with the
mass of the $D_s(2317)$.
As discussed by UKQCD~\cite{Green:2003zz}
a hadronic state will couple to many different 
interpolating
operators made out of quarks, anti-quarks and glue, with
the same quantum numbers as the hadron. 
In order to say whether a hadronic state is
more like a molecule
or $\overline{q}q$ state, the lattice calculation needs to also determine the
amplitude for a hadronic state to be in a 
specific configuration of quarks
and anti-quarks.
One possible way to determine the quark distribution 
of a hadronic state is to look
at observables like form factors,  the most basic being the decay constant.
For example, in a simple picture of a molecular
state, the decay constant should be suppressed relative to that of a
bound state of quark and antiquark.
An analysis of non-leptonic decays of the $B$ meson using the factorisation
hypothesis suggested that the decay constant of the $D_s(2317)$ could
be significantly smaller than that of the pseudo-scalar heavy-light
meson~\cite{Datta:2003re,Hwang:2004kg}. This motivates a lattice
calculation of the decay constant of the $0^+$ heavy-light meson.

The $B \to D^{\ast\ast} \pi$ decays are also relevant to clarify the so 
called `$1/2$ vs. $3/2$ 
puzzle'~\cite{Uraltsev:2004ra,Jugeau:2005yr,Bigi:2005ff}
Indeed, the heavy to heavy 
$B \to D^{\ast\ast}_j$ transitions, with $j=1/2,\ 3/2$, are parametrised 
by the generalised Isgur-Wise functions $\tau_j(w)$, where $w=v_B \cdot 
v_{D^{\ast\ast}_j}$. Several theoretical considerations using independent 
methods (sum rules derived from QCD~\cite{Uraltsev:2000ce}, 
covariant quark model~\cite{LeYaouanc:1995wv},
lattice QCD~\cite{Becirevic:2004ta,Blossier:2005vy},
 experimental data combined with naive 
factorisation~\cite{Jugeau:2005yr}), suggest that the $j=3/2$ states dominate, 
with respect to the $j=1/2$ states in $B \to D^{\ast\ast}$ transitions, 
i.e. $\tau_{3/2}(1) > \tau_{1/2}(1)$. Nevertheless this prediction is in 
contradiction with some of the currently available experimental data. The 
solution for this puzzle still requires new inputs from both theory and 
experiment.
From the theoretical side, the determination of the decay constant, 
$f_{D_{0^+}}$ of the $D_{0^+}$ meson is needed to evaluate the decay width 
of two (out of the three) classes of $B \to D^{\ast\ast} \pi$ decays, 
when naive factorisation is assumed~\cite{Jugeau:2005yr}. 
%%In the static limit, 
%%the decay constant, noted $f_{B_{1/2}}^{stat}$, of the degenerate 
%%$B_{j=1/2}$ mesons is a non-vanishing quantity, contrary to what happens 
%%in the case of the j=3/2 doublet~\cite{LeYaouanc:1996bd}.

A recent summary~\cite{Jugeau:2005yr} of results for the 
  decay constant,  showed that estimates for $f_{D_{0^+}}$
lay between 122 and 417 MeV. 
This summary
also reported that the decay constant of $0^+$ heavy-light
mesons in the static limit was higher than 
at the charm mass. This required confirmation,
because the heavy quark mass dependence was obscured 
by model dependence.
Some factorisation schemes~\cite{Dugan:1990de} 
for non-leptonic
decays of $B$ mesons use the static limit
as the leading term, hence computing the
decay constant of the P-wave static-light
meson is also important for this reason.
The evaluation of the $1/m_Q$ corrections to the heavy quark 
limit of the various quantities describing $B \to D^{\ast\ast}_j$ 
transitions is crucial since simple estimates tend to suggest the presence 
of large corrections~\cite{Jugeau:2005yr}.

%% FB, fBs
There is a huge effort in lattice
gauge theory that aims to compute the 
decay constant of the heavy-light pseudo-scalar mesons,
because they are crucial for determining CKM 
matrix elements~\cite{Lubicz:2004nn,Okamoto:2005zg}.
The decay constant of the heavy-light vector meson
has been computed in~\cite{Bernard:2001fz}
in order to test the
HQET scaling relations. 

There has been no published work
(apart from a private communication reported in~\cite{Jugeau:2005yr})
on using lattice QCD to compute the 
decay constants of P-wave heavy-light mesons.
The study of the
wave-function of P-wave static-light mesons~\cite{Duncan:1992eb},
using quenched lattice QCD, 
did not extract a decay constant.
UKQCD has studied the charge and matter radial distributions
of P-wave static-light mesons~\cite{Green:2004nm,Green:2005st}.

%%
%% data
%%

In our calculations,
we used the non-perturbatively 
improved clover action with clover
coefficient of 2.0171 for the light quarks. 
The Wilson gauge action was used.
The lattice volume was 
$16^3 \; 32$ and $\beta = 5.2$. There were two flavours
of sea quarks. The $\kappa$ values are tabulated
with the results.
The sea quark masses span from the strange quark
to one third of the strange quark mass. The details of 
the light hadron spectroscopy can be found 
in~\cite{Allton:2001sk,Allton:2004qq}.

The plan of the paper is 
that in section~\ref{se:prop}, we first
discuss our results using clover fermions 
with quark masses around the charm value.
Next in section~\ref{se:static},
we 
present data using static heavy quarks.
In the final section~\ref{se:concl} we compare
our results to other determinations and 
make some remarks about the heavy quark mass
dependence of the decay constants.

\section{Decay constants around the charm mass} \label{se:prop}

We have used the unquenched 
data from~\cite{Dougall:2003hv,Dougall:2005ev}
to study the decay constant of the $0^+$ meson around
the charm mass with clover quarks. 

%% data set
 The data set used the sea $\kappa$ = 0.1350
and three valence light quark masses using $\kappa$-values of
0.1340, 0.1345 and 0.1350. The $\kappa$
values for the heavy quarks were: 0.113,
0.119 and 0.125. Using $r_0$ = 0.5 fm, the inverse
lattice spacing is 1.88 GeV~\cite{Allton:2001sk}.
The sea quark mass was kept fixed at the mass of the strange
quark, hence this is a 
partially quenched calculation.
The data sample included 
394 gauge configurations separated by 20
trajectories. Adjacent data were binned together.
A smearing matrix of order two, using local and
fuzzed basis states, was fitted to a factorising fit
model~\cite{Allton:2001sk}.

%% definition of the decay constant
 The  decay constant of the $D_{0^+}$ meson
can be defined by
equation,
 \begin{equation}
\langle 0 \mid V_\mu^{cq} | D_{0^+} \rangle
= i p_\mu g_{0+}
\label{eq:VectorDefn}
 \end{equation}
where $V_\mu^{cq}$ is the vector current with two non-degenerate
quark flavours. 
 This uses the $V-A$ structure of the charged weak 
quark current
and parity conservation. We used a unit gamma
matrix between the heavy and light quark fields
as the interpolating operator for the $D_{0^+}$ state.
Unfortunately, we did not measure the lattice
correlators with  $\gamma_0$ at the sink and
$1$ at the source that would be required 
to use equation~\ref{eq:VectorDefn} directly.
In principle due to the number of gamma matrices,
there are 256 different local meson correlators.
Many lattice QCD codes do not compute every possible
correlator, because some correlators are zero
by symmetries. The correlator with a $\gamma_0$
at the sink and a unit matrix at the source
is zero for degenerate 
quarks because of charge conjugation.

Colangelo et al.~\cite{Colangelo:1991ug}
define the decay constant ($g_{0+}$) using 
equation~\ref{eq:decayDEFNCOL}.
 \begin{equation}
\langle 0 \mid \overline{c} q | D_{0^+} \rangle
= \frac{M_{D_{0^+}}^2}{m_c} g_{0+}
\label{eq:decayDEFNCOL}
 \end{equation}
%%%
 where $m_c$ is the mass of the charm quark.  
Equation~\ref{eq:decayDEFNCOL}  can be related to
equation~\ref{eq:VectorDefn} since the divergence 
of the vector current for non-degenerate quarks
is proportional to the scalar density. The derivation
of equation~\ref{eq:decayDEFNCOL} from 
equation~\ref{eq:VectorDefn}
assumes that the
mass of the light quark can be neglected
relative to that of the
charm quark.

From a lattice QCD perspective, it is more natural to
define the decay constant using
equation~\ref{eq:decayDEFN}.  
 \begin{equation}
\langle 0 \mid \overline{c} q | D_{0^+} \rangle
= M_{D_{0^+}} f_{0+}
\label{eq:decayDEFN}
 \end{equation}
 The determination of the mass of the 
charm quark can have quite large systematic errors
at a fixed lattice spacing~\cite{Dougall:2005ev}, so we
prefer the definition of the decay constant with
no explicit factor of the charm mass.
To convert from our normalisation of the decay constant $f_{0+}$ to
that of $g_{0+}$, 
we note that 
for this data set UKQCD obtained $m_c(m_c)^{\overline{MS}}$
= $1.247(3)^{+20}_{-4}$ GeV~\cite{Dougall:2005ev}, using
a quark mass defined using the Fermilab heavy quark 
formalism~\cite{El-Khadra:1996mp}.

The matrix element is related to the coupling amplitude $Z_i$
in the fit to the $0^+$ to $0^+$ meson correlator.
 \begin{equation}
Z_i = \frac{\langle 0 \mid \overline{c} q | D_{0^+}\rangle }
{ \sqrt{ 2 M_{D_{0^+}} } }
 \end{equation}
%%%
A simple linear fit model was used to 
extrapolate and interpolate in the quark masses.
We first interpolated to the strange
quark mass~\cite{Hepburn:2002wa}, 
or extrapolated to the light quark
mass (at $\kappa_{crit}$). The decay
constants were then interpolated to the 
value of the mass 
of the charm quark~\cite{Dougall:2005ev}.

\begin{figure}
\begin{center}
\includegraphics[scale=0.3,angle=270]{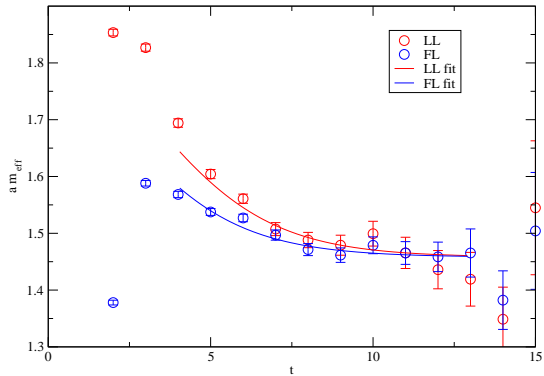}
\end{center}
\caption {
Effective mass plot and fitted model for the correlators with heavy 
$\kappa$ = 0.113 and light $\kappa$ = 0.1350.
}
\label{fig:meff}
\end{figure}

\begin{figure}
\begin{center}
\includegraphics[scale=0.3,angle=270]{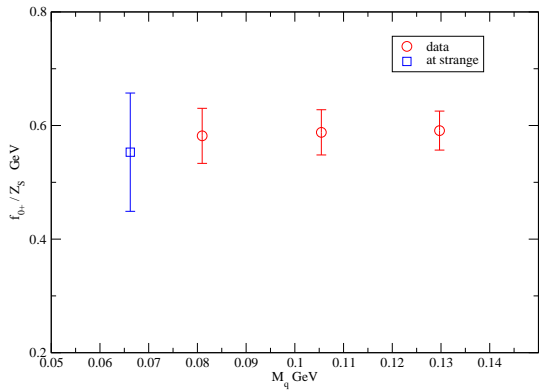}
\end{center}
\caption {
Mass dependence of the P-wave decay constant
defined in equation~\ref{eq:decayDEFN} at the 
fixed heavy quark mass with $\kappa$ = 0.113, as a function of the
bare light quark mass in physical units. 
The estimate of the decay constant at the mass of the strange 
quark is also shown.
}
\label{fig:MassDependlight}
\end{figure}

\begin{figure}
\begin{center}
\includegraphics[scale=0.3,angle=270]{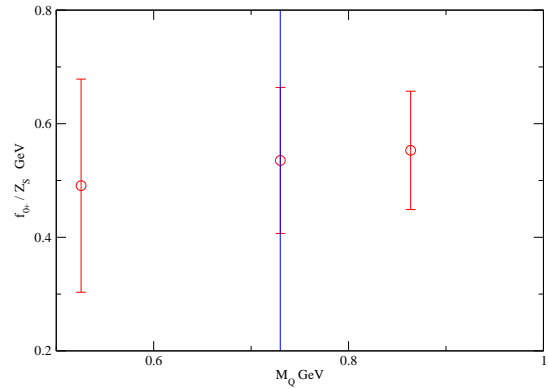}
\end{center}
\caption {
Mass dependence of the P-wave decay constant
defined in equation~\ref{eq:decayDEFN} at the 
strange quark mass, as a function of the
bare heavy quark mass in physical units. 
The vertical line is the estimate for
the charm quark mass.
}
\label{fig:MassDepend}
\end{figure}

In figure~\ref{fig:meff} we report an effective mass plot for the local-local and fuzzed-local
correlators used in the analysis.
In figure~\ref{fig:MassDependlight} we plot the bare decay
constant as a function of the bare light quark mass in
physical units with mass of the heavy quark fixed.
Similarly, figure~\ref{fig:MassDepend} shows the 
bare lattice decay constant as a function of the heavy
quark mass, with the light quark mass fixed at the 
mass of the strange quark.
The dependence of the decay constant on the mass of the
quarks is very mild. We discuss some of the issues 
in the chiral extrapolations in section~\ref{se:static}.

To convert the lattice number into the $\overline{MS}$ 
scheme we use 
tadpole improved perturbation theory to one
loop order (the required expressions
are listed in~\cite{Gockeler:1996gu,Bhattacharya:2000pn}).

\begin{equation}
Z_S = u_0  
\left( 
1 + \alpha_s ( \frac{1}{\pi} \log(\mu a)^2 -1.002  ) 
\right)
\label{eq:ZStad}
\end{equation}
 where $u_0$ is the fourth root of the plaquette.
The mass dependent  improvement factor
$(1 + b_S \frac{m_{i} + m_{j}}{2})$
was multiplied into the scalar current.
We used the one loop expression for 
$b_S$~\cite{Sint:1997dj},
\begin{equation}
b_S = \left( 1 + \alpha_s 1.3722  \right) 
\end{equation}
and the same prescription for the coupling as used
in UKQCD's estimate of the mass of the charm 
quark~\cite{Dougall:2005ev}. We use $\mu=a^{-1}$
to determine $Z_S$.

In lattice units we obtain $f_{D_{0^+}} / Z_S$ =  0.280(70)
and $f_{D_{s\,0^+}} / Z_S$ = 0.270(90).
Using $r_0$ = 0.5 fm,
and $Z_S$ from equation~\ref{eq:ZStad}, we obtain
$f_{D_{0^+}}$ =  360(90) MeV
and $f_{D_{s\,0^+}}$ = 340(110) MeV.
In principle,
we should estimate the effect of the
lattice spacing dependence and higher order terms
in the perturbative expansion, as we did when we 
estimated the mass of the bottom quark~\cite{McNeile:2004cb}.
However, because the statistical errors are of the order
of 30\%, the statistical errors will clearly dominate.
For example, a typical estimate of the systematic error due to
the different ways of determining the lattice
spacing in a lattice calculation with these parameters would be to
vary the value of $r_0$ between 0.5 fm and 0.55 fm. This systematic
error is a 10 \% effect.

We now discuss the central value of $r_0$ used.  When $r_0$, a
number derived from the heavy quark potential, was introduced by
Sommer~\cite{Sommer:1993ce}, 
he quoted a value of $r_0$ about $0.49$ fm. The value of
$r_0$ can also be determined from lattice calculations. As reviewed
in~\cite{Dougall:2003hv}, unquenched calculations with similar
parameters to those used here compute a value of $r_0$ between
0.5 and 0.55 fm. The issue is not so clear because the 
HPQCD collaboration, using gauge configurations from the
MILC collaboration, obtain a value of $r_0$ = 0.469(7) fm.
As UKQCD has argued~\cite{McNeile:2004cb}, 
for calculations with these parameters it better to use a
value of $r_0$ closer to 0.5 fm because this is the value
determined from similar calculations. The MILC/HPQCD value for 
$r_0$ is with in 10\% of 0.5 fm.

\section{Decay constants in the static limit} \label{se:static}

This part of the work is a continuation of UKQCD's study of static-light
mesons. The static-light meson spectrum was reported
in~\cite{Green:2003zz}, and then used  to
extract the mass of the bottom quark~\cite{McNeile:2004cb}.
Here we use the extended data set that was used
to look for chiral logs in the $f_B$ decay 
constant~\cite{McNeile:2004wn}.

In addition to using the standard static action
of Eichten and Hill,
we also report data using  one of the 
static actions developed by the ALPHA 
collaboration~\cite{DellaMorte:2003mn}
with improved signal to noise ratio.
The exact static actions are described in~\cite{McNeile:2004wn}.

Here we focus on the meson doublet with total light quark momentum
of $j=1/2$, we call this the
$P_-$ static-light meson~\cite{Green:2003zz}. 
As 
corrections to the static limit are
included the $P_-$ state will split
into a $0^+$ state and a $1^{+}$ state. In this paper, the decay
constant of the $P_-$ static-light meson
is called $f_P^{static}$.
 %%
%% definition of the decay constant
%% 
 The $f_P^{static}$ decay constant is defined by the matrix element 
in equation
 \begin{equation}
\langle 0 \mid V_\mu \mid P_{-}(p) \rangle = i p_\mu f_P^{static}
\label{eq:DECAYdefn}
\end{equation}
 %%%%
 where $V_\mu$ is the vector 
current~\cite{Colangelo:1991ug,Colangelo:1992kc}.
The simple spin structure of the static quark means
the same decay constant is obtained if
in equation~\ref{eq:DECAYdefn} the vector current  (with $\mu=0$)
is replaced with a 
scalar current.
 %%
%%
%% defintion of the pole
 %%
The $f_P^{static}$ matrix element is extracted from 
the amplitudes in the two point correlator
 %%%%
 \begin{eqnarray}
C(t) & = &  \sum_{x} 
\langle 0 \mid 
V_0(x,t)
\Psi_B^\dagger(x,0) 
\mid 0 \rangle \\
     & \rightarrow & Z_L^{static} Z_{\Psi_B} \exp( - a{ \cal E } t)
\; ,
\label{eq:fitModel}
 \end{eqnarray}
 where  $\Psi_B$ is the interpolating operator for  
the $P_-$
static-light meson
and ground state dominance is shown in equation~\ref{eq:fitModel}. 
In practise we use a basis of smearing functions 
to do variational smearing~\cite{Green:2003zz}.
The
$Z_{L}^{static}$ amplitude is related to the $f_P^{static}$ decay constant
 \begin{equation}
f_P^{static} = Z_{L}^{static} \sqrt{ \frac{2}{M_{B_{0^+}} }} Z_S^{static}
\label{eq:ZLstaticamp}
 \end{equation}
 where $Z_S^{static}$ is a perturbative matching
factor that we discuss below.
This is an equivalent definition to the one
used for the pseudo-scalar heavy-light decay
constant ($f_B^{static}$).

%%
%% renormalisation issues
%%
 It is traditional to match the results of a lattice static-light
calculation to continuum QCD via two steps~\cite{Eichten:1989zv}. 
QCD is matched to the continuum static theory. The lattice
static theory is then matched to the static continuum 
theory. 

The matching of the static continuum theory 
to the static-lattice theory has been done
by Eichten and Hill~\cite{Eichten:1990kb}  
and by Borrelli and Pittori~\cite{Borrelli:1992fy} for
the clover action. This matching was done for 
a heavy-light current with arbitrary 
gamma matrix ($\Gamma$). For the definition of the
decay constant in equation~\ref{eq:DECAYdefn}, we
do the perturbative matching assuming $\Gamma=\gamma_0$.
The matching between the continuum static theory
and the static lattice theory is via $Z(\Gamma)$.
\begin{equation}
Z(\Gamma) = 1 + \frac{g^2}{12  \pi^2}    \left( 
\frac{3}{2} \log( \mu^2  a^2) +
5/4 - A_{\Gamma}  \right)
\end{equation}
where
\begin{equation}
A_{\Gamma} = d_1 + (d_2 - d^1)G + \frac{(e  + f + f^1 )}{2}
\end{equation}
where the values of the constants are
$d_1 = 5.46$,
$d_2 = -7.22$,
$f = 13.35$, $e=4.53$~\cite{Eichten:1990kb},
$d^1 = -4.04$, and  $f^1 = -3.63$~\cite{Borrelli:1992fy}.
$G=1$ when $\Gamma = \gamma_0$.

The matching of the continuum static theory to
continuum QCD was done using~\cite{Borrelli:1992fy}
\begin{equation}
Z_{Qstat} = 1 + \frac{g^2}{ 12 \pi^2} \left( -\frac{3}{2} \log(\mu^2/m_b^2) -2  \right)
\; .
\label{eq:ZcontStat}
\end{equation}
where $m_b$ is the mass of the bottom quark.
The equivalent expression for equation~\ref{eq:ZcontStat},
for arbitrary $\Gamma$ matrix,
is in~\cite{Borrelli:1992fy}. 
The matching factor $Z_S^{static}$ in equation is the product 
of $Z(\Gamma=\gamma_0)$ with $Z_{Qstat}$ to leading order
in the square of the coupling $g^2$.

The improvement coefficients have not been
calculated for the P-wave decay constant and we will therefore not
include them
in our analysis.
We used a simple boosted coupling ($g^2/(u_0)^4$ and 
$g^2 = \frac{6}{\beta}$) to compute the 
perturbative matching factors.

We use the mass ($5279+400$ MeV) 
for the mass factor in
equation~\ref{eq:ZLstaticamp}. The experimental spectrum
of the P-wave B mesons is not very well determined at the 
moment~\cite{Swanson:2006st,Eidelman:2004wy}. Lattice
calculations predict that the lowest P-wave meson will be 
roughly 400 MeV above the S-wave states. 
So we add 400 MeV to the mass of the $B^+$ state~\cite{Eidelman:2004wy}.
This corresponds to a 4\% effect
on the decay constant.
We prefer not to use  the HQET
scaling law of the decay constant to quote numbers for the decay
constant of the $0^{+}$ charm-light meson, as is done by Jugeau et
al.~\cite{Jugeau:2005yr}, because it is known that $1/M$ corrections to
the static limit of the decay constant of the pseudo-scalar
heavy-light meson are large.

The unrenormalised data for the amplitudes are
in table~\ref{tab:STATICbareRESULTS}. The final
column shows the value for $f_P^{static}$ in the static
limit in physical units. The value of $r_0$ = 0.5 fm
was used to convert the data into physical
units. 

As  the renormalisation factor for
the  ALPHA static action has not yet been determined,  we can't use 
that data to quote a physical number.
What is disappointing
is that the ALPHA static action does not produce
any reduction in the statistical errors over the
standard Eichten-Hill static action in our case. For the 
static-light 
pseudo-scalar meson decay constant the
statistical errors were significantly smaller
for the ALPHA static action than the Eichten-Hill
action~\cite{McNeile:2004wn}. 

In~\cite{Green:2003zz} it was shown that the 
$DF3$ data set corresponded to sea quarks
with mass around the strange quark mass.
Hence we quote the $f_{P_s}^{static}$ decay 
constant as $302(39)$ MeV.

%%
%% table of data
%%
\begin{table}[tb]
\begin{center}
\begin{tabular}{|c|ccc|c|} \hline
Name & formalism  & $\kappa_{sea}$ & $Z_{L}^{static}$       & $f_P^{static}$ MeV   \\ \hline
DF3 &  Eichten-Hill & 0.1350 &  $0.240_{-27}^{+31}$ & 302(39)   \\
DF3 &  Fuzzed ALPHA & 0.1350 &  $0.199_{-20}^{+35}$ &   \\
DF4 &  Eichten-Hill & 0.1355 &  $0.234_{-18}^{+43}$ & 322(59)  \\
DF4 &  Fuzzed ALPHA & 0.1355 &  $0.174_{-20}^{+30}$ & \\ 
%%%
DF6 &  Eichten-Hill & 0.1358 &  $0.182_{-36}^{+52}$ &  272(78)\\
DF6 &  Fuzzed ALPHA & 0.1358 &  $0.131_{-35}^{+41}$ &  \\ \hline 
%%% draper_magic  \vsbtc
\end{tabular}
  \caption{
Amplitudes and decay constants for the $P_{-}$ static-light mesons.
Further information about the lattice parameters is in~\cite{McNeile:2004wn}.
}
\end{center}
\label{tab:STATICbareRESULTS}
\end{table}

The value of the 
P-wave $0^{+}$ decay constant at the
strange quark mass is of interest to compare to
model calculations or for future decays of 
the $B_s$ meson measured at a hadronic experiment
such as LHCb, CDF, or D0.
The decay constant required for 
the factorisation analysis of the heavy to heavy
$\overline{B} \rightarrow D^{\star \star} \pi$
is the $f_P^{static}$  decay constant 
in the chiral limit~\cite{Jugeau:2005yr}. 

To study the light P-wave $0^+$ decay constant 
the chiral extrapolation must be discussed.
The importance of chiral logs in the mass 
extrapolation of the $f_B$ decay constant has
only recently been observed as the dominant systematic error
in the determination of $f_B / f_{B_s}$~\cite{Aoki:2003xb,Kronfeld:2002ab}. 

The equivalent chiral perturbation theory calculation
for the P-wave decay constant (to our knowledge) 
has not yet been done. There are calculations of the 
masses of heavy-light mesons to one loop
in heavy-light chiral perturbation 
theory~\cite{Becirevic:2004uv,Mehen:2005hc}.
UKQCD has previously computed the relevant 
hadronic coupling~\cite{McNeile:2004rf}. This result is compared
against experiment in~\cite{Becirevic:2004uv}.

The data for the P-wave decay constant 
in table~\ref{tab:STATICbareRESULTS} 
are not really precise enough to determine
the quark mass dependence. 
A simple linear fit against the square of the pion mass of the data in
table~\ref{tab:STATICbareRESULTS} gives $f_P^{static} = 294(88)$ MeV
in the chiral limit of zero mass light quarks.

\section{Conclusion} \label{se:concl}

We have computed the decay constant of 
$0^+$ heavy-light mesons using an unquenched
lattice QCD calculation at a single lattice spacing.
For the static-light decay constant, we used 
sea quarks as low at a third of the strange quark mass.
For the calculation of the $0^+$ decay constant,
we did a partially quenched analysis with the 
sea quark fixed at the strange quark mass.

We obtain the $f_{P_s}^{static}$ decay
constant as $302(39)$ MeV. Given the qualifications
mentioned in the previous section, we obtain
$f_{P}^{static} = 294(88)$ MeV. The data for
$f_{P}^{static}$ from various models 
are summarised in~\cite{Jugeau:2005yr}.
Two QCD sum rule results are $f_{P}^{static}$
=$304 \pm 40$~\cite{Colangelo:1992kc} 
and $377 \pm 53$ MeV~\cite{Zhu:1998wy}.
%%
%%  1/M corrections
%%
 Using clover quarks for the charm quark, we obtain
$f_{D_{0^+}}$ =  360(90) MeV
and $f_{D_{s\,0^+}}$ = 340(110) MeV.
The magnitude of these decay constants can 
be compared against the value of the $f_{D_s}$ and 
$f_{D}$ pseudo-scalar
decay
constants. From unquenched lattice QCD, the Fermilab, MILC and
HPQCD collaborations~\cite{Aubin:2005ar} 
obtained:
$f_{D_s} = 249 \pm 3 \pm 16$ MeV
and
$f_{D^+} = 201 \pm 3 \pm 17$ MeV.
The CLEO-c collaboration 
have recently reported~\cite{Artuso:2005ym}
the experimental result  
$f_{D^+} = 223(17)(3)$ MeV.

Although the systematic errors are different in the 
two lattice calculations,
we can make the qualitative statement that 
the P-wave and S-wave decay constants of charm-light mesons are
similar in magnitude. This calculation does not support
a suppression of the decay constant of the P-wave heavy-light meson
relative to the decay constant of the S-wave 
heavy-light meson.

The only decay constant of a light P-wave meson
that is sometimes calculated from lattice QCD is that of the $a_1$
meson. The decay constant of the $a_1$ meson, denoted as $f_{a_1}$,
is usually computed from a definition that has
dimension $\mbox{MeV}^2$. If we use the results from
Wingate et al.~\cite{Wingate:1995hy}, then the value of
$\frac{f_{a_1}}{M_{a_1}}$ is 240 MeV. This is larger
than the value of the pion decay constant 131 MeV, showing
a similar trend to the heavy-light case.

\begin{table}
\begin{center}
\begin{tabular}{|c|c|}
\hline 
Method & $g_{0+}$  MeV \\ \hline
This work & $200 \pm 50$ \\
QCD sum rules~\cite{Colangelo:1991ug} & $170  \pm 20$ \\
Lattice QCD~\cite{Jugeau:2005yr} & $122 \pm 43$ \\
$B \rightarrow D^{\star \star}\pi$~\cite{Jugeau:2005yr}  & $206 \pm 120$ \\ \hline
\end{tabular}
\label{ZZZ}
  \caption{
Comparison of decay constants of charm-light
$D_{0^+}$ meson.
}
\end{center}
\end{table}

Kurth and Sommer~\cite{Kurth:2001yr} have noted that there are
potential problems with extrapolating in the heavy-quark mass unless
the continuum limit has been taken. 
Also see Kronfeld~\cite{Kronfeld:2002pi} for a discussion
of the problems combining static and propagating heavy-light
data.
Given these theoretical
concerns, the data is consistent with the view that the decay
constant of the static-light meson is larger than that
of the $0^{+}$ charm-light meson, if we use definition of the
decay constant in equation~\ref{eq:VectorDefn} (our value of 
$g_0$ is in table 2),
%%$g_0$ is in table~\ref{ZZZ}),
but the errors need to be
reduced for a definitive statement.

Jugeau et al.~\cite{Jugeau:2005yr}. have collected
together a number of different calculations of 
the $0^{+}$ decay constant of a charm-light meson.
To compare our results to other calculations we
use the normalisation in equation~\ref{eq:VectorDefn}.
Hence we multiply our result by 
$m_c/M_{D_s(2317)}$ = 1.27/2.317 and
compare to other calculations in 
table 2. 
%%table~\ref{ZZZ}. 
There is reasonable
agreement between different determinations. To reduce 
the errors on the lattice results requires 
a calculation of similar effort to that done to compute
the decay constants of the
heavy-light pseudo-scalar mesons~\cite{Aubin:2005ar}.
%%%
%%%

%%
%%
%%

\section{Acknowledgements}

We thank Alex Dougall for help.
The gauge configurations were generated on the Cray T3E system at EPCC
supported by EPSRC grant GR/K41663, PPARC grants GR/L22744 and
PPA/G/S/1998/00777.  This work has been supported in part by the EU Integrated
 Infrastructure Initiative Hadron Physics (I3HP) under contract
  RII3-CT-2004-506078. We are grateful to the ULgrid project of the
University of Liverpool for computer time. One of the authors (CM)
wishes to thank PPARC for the award of a Senior Fellowship.

%%\bibliographystyle{h-physrev2}
%%\bibliography{q_mass} 

\end{document}